\documentclass[preprintnumbers, aps, floatfix, onecolumn, preprintnumbers, letterpaper, superscriptaddress,nofootinbib,10pt]{revtex4-2}
\usepackage{dcolumn}
\usepackage{graphicx}
\usepackage{amsmath}
\usepackage{amsfonts}
\usepackage{amssymb}
\usepackage{microtype}
\usepackage{subfigure}
\usepackage{makeidx}
\usepackage{bm}
\usepackage{epsf}
\usepackage[colorlinks=true,citecolor=blue,linkcolor=black,urlcolor=black]{hyperref}
\usepackage{color}
\usepackage{multirow,dcolumn}
\usepackage{graphicx}
\usepackage{mathrsfs}
\graphicspath{{Images/}}

\def\doi{http://doi.org}

\def\be{\begin{equation*}}
\def\ee{\end{equation*}}

\def\Ref{\ref}

\begin{document}

\title{$f({\sf R})$ Gravity Wormholes sourced by a Phantom Scalar Field}

\author{Thanasis Karakasis}
\email{thanasiskarakasis@mail.ntua.gr} \affiliation{Physics Division,
National Technical University of Athens, 15780 Zografou Campus,
Athens, Greece.}

\author{Eleftherios Papantonopoulos}
\email{lpapa@central.ntua.gr} \affiliation{Physics Division,
National Technical University of Athens, 15780 Zografou Campus,
Athens, Greece.}

\author{Christoforos Vlachos}
\email{cvlach@mail.ntua.gr} \affiliation{Physics Division,
National Technical University of Athens, 15780 Zografou Campus,
Athens, Greece.}

\begin{abstract}
We derive an exact wormhole spacetime supported by a phantom scalar field in the context of $f({\sf R})$ gravity theory. Without specifying the form of the $f({\sf R})$ function, the scalar field self-interacts with a mass term potential which is derived from the scalar equation and in the resulting $f({\sf R})$ model  the scalar curvature is modified by the presence of the scalar field and it  is free of ghosts and  avoids the tachyonic instability.

\end{abstract}

\maketitle


\tableofcontents

\section{Introduction}

Wormholes in General Relativity (GR) are solutions of Einstein equations presenting hypothetical tunnels connecting different parts of the Universe or two different Universes. The first discussion of a wormhole configuration was presented by Flamm \cite{Flamm} and later by Einstein and Rosen (ER) introducing the “ER bridge”,  a physical space being connected by a wormhole-type solution \cite{Rosen}. The wormhole spacetimes as solutions of GR were further investigated in the  pioneering articles of Misner and Wheeler \cite{MisWheel} and Wheeler \cite{Wheel}.

Introducing a  static spherically symmetric metric the necessary conditions to generate traversable Lorentzian wormholes as  exact solutions in GR were first found  by Morris and Thorne \cite{Morris}. In these structures a condition on the wormhole throat necessitates the introduction of   exotic matter, which however  leads to the violation of the null energy condition (NEC). This type of matter has been  discussed in cosmological contexts \cite{phantom}, for possible  observational applications. Recently, it was found in \cite{Konoplya:2021hsm} that normal Dirac and Maxwell fields can support wormhole configurations and provide violation only of the NEC. Many wormhole solutions were discussed in the literature. To avoid the violation  of NEC in \cite{Visser}  the construction of thin-shell wormholes was  studied, where ordinary matter is concentrated   on the wormhole throat.
Recently, there are many studies of wormhole solutions in modified gravity theories like Brans-Dicke theory \cite{Brans}, mimetic theories \cite{Myrzakulov:2015kda}, $f({\sf R})$ gravity \cite{FR},  Einstein-Gauss-Bonnet theory~\cite{GB}, Einstein-Cartan theory and general scalar-tensor theories \cite{Cartan}.
 Wormholes with AdS asymptotics have also been discussed \cite{Maldacena:2004rf} in an attempt to describe the physics of closed universes. Canonical scalar fields in the Horndeski scenario have also been used to constract wormhole geometries \cite{Korolev:2014hwa,Korolev:2020ohi,Korolev:2020yyy}. In these theories, their gravitational echoes have been recently investigated \cite{Vlachos:2021weq}, while the formation of bound states of scalar fields in AdS-asymptotic wormholes were studied in \cite{Chatzifotis:2020oqr}. A phantom scalar field was introduced in the Einstein-Hilbert action and wormhole solutions have been found by Ellis in \cite{Ellis:1973yv}, known as Ellis wormholes. Lately, several generalizations were obtained \cite{Blazquez-Salcedo:2020nsa, Chew:2016epf, Chew:2018vjp, Blazquez-Salcedo:2018ipc, Sahoo:2018kct}. Phantom matter was considered in Einstein-Maxwell dilaton theory and wormholes have been constructed in \cite{Goulart:2017iko}.

The need to describe the early and late cosmological history of our Universe  promoted the study of $f({\sf R})$ theories of gravity
\cite{DeFelice:2010aj, Cognola:2007zu, Zhang:2005vt, Li:2007xn, Song:2007da, Nojiri:2007cq, Nojiri:2007as, Capozziello:2018ddp, Starobinsky:1980te}. The main motivation to study these theories were  the  recent cosmological observational results on the  deceleration-acceleration transition of late Universe. This requirement imposed constraints on  the $f({\sf R})$ models allowing viable choices of $f({\sf R})$ \cite{Capozziello:2014zda}.
These theories  exclude contributions from any curvature invariants other than $\sf R$ and they avoid the Ostrogradski instability \cite{Ostrogradsky:1850fid,Woodard:2006nt}.

Black holes in  $f({\sf R})$ gravity theories  with constant and non-constant Ricci curvature in vaccum or coupled to electrodynamics have been found
\cite{Sebastiani:2010kv}- \cite{Tang:2019qiy}
while in \cite{Tang:2020sjs,Karakasis:2021lnq,Karakasis:2021rpn} scalar fields are introduced as matter in $(2+1)$ and $(3+1)$-dimensional $f({\sf R})$ gravity and the corresponding black hole solutions are investigated. This particular type of theory, $f({\sf R})$ gravity and non-minimally coupled scalar fields as matter has been previously considered for cosmological purposes \cite{Pi:2017gih, delaCruz-Dombriz:2016bjj}.

In \cite{Karakasis:2021lnq} an exact black hole solution in $(2+1)$-dimensions of a scalar field minimally coupled to gravity in the context of $f({\sf R})$ gravity was found.  Without specifying the form of the $f({\sf R})$ function, an exact black hole solution was obtained dressed with a scalar hair and the scalar charge to appear both in the metric  and in the $f({\sf R})$ function. It was showed that thermodynamical and observational constraints required that the pure $f({\sf R})$ theory should be builded with a phantom scalar field. The reason for this behaviour is that the entropy in the $f({\sf R})$ theories receives an extra contribution and to have a positive entropy a contribution from  phantom scalar field is required. Then, computing  the Hawking temperature and the Bekenstein-Hawking entropy it was found that they are both positive, with the temperature getting smaller with the increase of the scalar charge  while the entropy behaves the opposite way, growing with the increase of the scalar charge.

 Wormhole solutions in $f({\sf R})$ theory have been constructed in \cite{Lobo:2009ip}, where the matter threading the wormhole is a fluid that satisfies the energy conditions, so the violation of the energy conditions comes from the higher order terms that the $f({\sf R})$ theory possesses. Thin-shell wormholes with circular symmetry in $(2+1)$-dimensional $f({\sf R})$ theories of gravity, with constant Ricci scalar have been investigated \cite{Bejarano:2021tau}, while $(3+1)$-dimensional thin-shell wormholes in $f({\sf R})$ gravity with charge and constant curvature were discussed in \cite{Eiroa:2015hrt}, investigating also their stability. In quadratic $f( \sf R)= \sf R+\alpha \sf R^2$ gravity spherically symmetric Lorentzian wormholes have been found with a constant scalar curvature \cite{Eiroa:2016zjx}. Using the Karmarkar condition wormhole geometries have been investigated in several $f({\sf R})$ models \cite{Shamir:2020uzy}. In \cite{Chervon:2020kdv} wormholes with a kinetic curvature scalar were considered and in \cite{Godani:2020vqe} traversable $f({\sf R})$ gravity wormholes with constant and dynamic redshift function were found. Finally in \cite{Tangphati:2020mir} wormholes in $f({\sf R})$-massive gravity were investigated employing several behaviors for the redshift function, and wormholes in generalized hybrid metric-Palatini gravity that satisty the NEC were obtained \cite{Rosa:2018jwp}.

  In
$f({\sf R})$ gravity theories  if  a conformal transformation is applied from the Jordan frame to the Einstein frame then, a new
scalar field appears and also a scalar potential is generated. The resulted theory can be considered as a scalar-tensor
theory with a geometric (gravitational) scalar field which however cannot dress a $ f({\sf R})$  black hole with hair \cite{Canate1,Sultana1,Canate2}. In our previous work \cite{Karakasis:2021lnq,Karakasis:2021rpn}
the motivation for constructing  hairy black hole solutions in  $f({\sf R})$ gravity theories was to introduce a scalar field  in the action as a matter field in the way it was done in the GR context
and study its effect on a metric ansatz solving the field equations. In these models the scalar field was a canonical scalar field with positive kinetic energy.

In this work we follow a similar approach. We would like to investigate  if we introduce a scalar field with negative kinetic energy (phantom scalar field)  with a self-interacting potential, a wormhole configuration can be generated. In the literature in most of the studied models, the phantom matter is introduced in the energy-momentum tensor with a phantom  equation of state violating the NEC.  In our study we introduce explicitly a self-interacting phantom field and  without specifying the form of the $f({\sf R})$ function, we solve the resulting field equations. To do that we have to specify the form of the phantom field and we find a new wormhole geometry sourced by the phantom scalar field which is free of ghosts and avoids the tachyonic instability and we show that the NEC is violated.

This work is organized as follows. In Section \ref{sect2} we set up our theory, derive the field equations and discuss the restrictions a geometry has to obey  to represent a wormhole configuration. In Section \ref{wormholes} we briefly discuss the GR case of a wormhole sourced by a phantom scalar field, the Ellis Drainhole \cite{Ellis:1973yv} and we obtain a new wormhole geometry sourced by a phantom scalar field in $f({\sf R})$ gravity discussing also the energy conditions. Finally, in Section \ref{conc} we conclude and we discuss possible extensions  for further work.

\section{The setup-derivation of the field equations}
\label{sect2}

We consider the following action
\begin{equation} S = \int d^4x \sqrt{-g} \left( \frac{f(R)}{2} + \frac{1}{2}\partial^{\mu}\phi\partial_{\mu}\phi -V(\phi) \right) ~,\label{action} \end{equation}
which consists of an arbitrary differentiable function of the Ricci Scalar $f(R)$, a scalar field with negative kinetic energy and a self-interacting potential.

The field equations that arise by variation of this action  are
\begin{eqnarray}
R_{\mu\nu}f_R(R) - \frac{1}{2}g_{\mu\nu}f(R) + \Big(g_{\mu\nu} \Box - \nabla_{\mu} \nabla_{\nu}\Big)f_R(R) &=& T_{\mu\nu}^{\phi}~, \label{EE}\\
\Box \phi + V_{\phi}(\phi) &=&0~, \label{KG}
\end{eqnarray}
where $\Box = \nabla^{\mu}\nabla_{\mu}$ is the D'Alambert operator with respect to the metric, $f_R(R) = df(R)/dR$ and the energy momentum tensor is given by
\begin{equation} T_{\mu\nu}^{\phi} = - \partial_{\mu}\phi \partial_{\nu}\phi + \frac{1}{2}g_{\mu\nu}\partial^{\alpha}\phi \partial_{\alpha}\phi - g_{\mu\nu}V(\phi)~. \end{equation}
We consider the following metric ansatz firstly used by  Morris and Thorne \cite{Morris} in spherical coordinates
\begin{equation} ds^2 = - e^{2\Phi(r)}dt^2 + \left(1-\frac{b(r)}{r}\right)^{-1} dr^2 + r^2 d\Omega\label{MorrisThorne-metric} ~,\end{equation}
where $\Phi(r)$ is the redshift function and $b(r)$ is the shape function and $d\Omega = d\theta^2 + \sin(\theta)^2d\varphi^2$. In order to obtain a wormhole geometry, these functions have to satisfy the following conditions \cite{Morris,Visser:1995cc}, namely:
\begin{enumerate}
    \item $\frac{b(r)}{r}\leq 1$ for every $[r_{th},+\infty)$, where $r_{th}$ is the radius of the throat. This condition ensures that the proper radial distance defined by $l(r)=\pm\int_{r_{th}}^{r}\left(1-\frac{b(r)}{r}\right)^{-1} dr$ is finite everywhere in spacetime.
   Note that in the coordinates $(t,l,\theta,\phi)$ the line element \eqref{MorrisThorne-metric} can be written as \begin{align}
    ds^2=-e^{2\Phi(l)}dt^2+dl^2+r^2(l)(d\theta^2+\sin^2\theta d\varphi)\,.\label{metric-proper-chart}
    \end{align}
   In this case the throat radius would be given by $r_{th}=\min\{r(l)\}$.
    \item $\frac{b(r_{th})}{r_{th}}=1$ at the throat. This relation comes from requiring the throat to be a stationary point of $r(l)$. Equivalently, one may arrive at this equation by demanding the embedded surface of the wormhole to be vertical at the throat.
    \item $b'(r)<\frac{b(r)}{r}$ which reduces to $b'(r_{th})\leq 1$ at the throat. This is known as the flare-out condition since it guarantees $r_{th}$ to be a minimum and not any other stationary point.
    \end{enumerate}

Moreover, to simplify the calculations and to ensure the absence of horizons we will consider the case $\Phi(r) =C$ as in \cite{Lobo:2009ip}, where $C$ is a constant. Computing the $tt,rr,\theta\theta$ and the Klein-Gordon equations we find
\begin{eqnarray}
&&r b' f'_R+3 b f'_R+2 r (b-r) f''_R-r b \phi '^2-4 r f'_R+r^2 f-2 r^2 V+r^2 \phi '^2=0 ~, \label{tt1} \\
&&2 f_R \left(b-r b'\right)+4 r (b-r) f'_R+r^2 (b-r) \phi '^2+r^3 (f-2 V)=0~ , \label{rr1} \\
&&r \left(r \left(\left(b'-2\right) f'_R+r \left(-2 f''_R+f-2 V+\phi '^2\right)\right)-f_R b'\right)-b \left(r \left(-f'_R-2 r f''_R+r \phi '^2\right)+f_R\right)=0 \label{uu1} \\
&&\frac{\left(r \left(b'-4\right)+3 b\right) \phi '+2 r (b-r) \phi ''}{2 r^2}-\frac{V'}{\phi '}=0~, \label{KG1}
\end{eqnarray}
where primes denote derivative with respect to the argument, the radial co-ordinate $r$ and $f=f(R)=f(r)$, the gravitational model as a function of the radial co-ordinate $r$. Eliminating $f$ from equation (\Ref{tt1}) we obtain
\begin{equation} f= 2 V-\frac{\left(r \left(b'-4\right)+3 b\right) f'_R+r (r-b) \left(\phi '^2-2 f''_R\right)}{r^2}~. \end{equation}
Now substituting $f$ back to equations (\Ref{rr1}) and (\Ref{uu1}) we can obtain the following equations
\begin{eqnarray}
&& r \left(r \left(b' f'_R-2 r f''_R+2 r \phi '^2\right)+2 f_R b'\right)-b \left(r \left(f'_R-2 r f''_R+2 r \phi '^2\right)+2 f_R\right) = 0~, \label{ttrr}\\
&& r \left(f_R b'-2 r f'_R\right)+b \left(2 r f'_R+f_R\right) = 0~. \label{ttuu}
\end{eqnarray}
As can be seen in equation (\ref{ttuu}), there is is a direct relation between the geometry and the $f(R)$ gravity,   while the scalar field will affect both the $f(R)$ function and the geometry as we can see in equation (\ref{ttrr}).

\section{Wormhole Solutions}

 In this Section we will discuss the wormhole solutions of the field equations. We will first review the Ellis Drainhole.
\label{wormholes}

\subsection{The Ellis Drainhole}

The Ellis Drainhole \cite{Ellis:1973yv} is a wormhole solution of an action that consists of a pure Einstein-Hilbert term and a scalar field with negative kinetic energy
\begin{equation}
	S = \int d^4x \sqrt{-g} \left(\frac{R}{2} + \frac{1}{2}\partial^{\mu}\phi\partial_{\mu}\phi\right)~.
\end{equation}
Assuming the metric ansatz (\ref{MorrisThorne-metric}), setting in equations (\Ref{tt1}),(\Ref{rr1}),(\Ref{uu1}) calculated above $f(r)=R(r), f_R(r) =1, V(r)=0$ we obtain the following solution
\begin{eqnarray}
b(r) &=&A^2/r~, \label{bellis}\\
\phi(r) &=& \sqrt{2} \tan ^{-1}\left(\frac{\sqrt{r^2-A^2}}{A}\right)+\phi_{\infty}~, \label{ellisscalar}\\
f(r) &=& R(r) =-\frac{2 A^2}{r^4}~, \label{rellis}
\end{eqnarray}
where $A, \phi_{\infty}$ are two constants of integration. The resulting spacetime is asymptotically flat as it can be seen from both $b(r)$ and $R(r)$. The wormhole throat is the solution of the equation
\begin{equation} g_{rr}^{-1} = 0 \rightarrow r_{th} = \pm A~. \end{equation}
The solution also satisfies the flaring-out condition and $b'(r_{th})=-1$.

 The scalar field takes a constant value at infinity $\phi(r\to \infty) = \frac{\pi }{\sqrt{2}} +\phi_{\infty} $, so one could set $\phi_{\infty} = -\frac{\pi }{\sqrt{2}}$ to make the scalar field vanish at large distances. However, since only derivatives of the scalar field appear in the field equations, its asymptotic value $\phi_{\infty}$ does not change the physical interpretation of the solution. The scalar field takes the asymptotic value at infinity at the position of the throat $\phi(r=r_{th})=\phi_{\infty}$. As can be seen in the solution (\ref{bellis}), (\ref{ellisscalar}) and (\ref{rellis}) the integration constant $A$  of the phantom scalar field plays a decisive role in the formation of the wormhole geometry.  It has  units $[L]$, appears in the scalar curvature and at the position of the throat takes the value $R_{r = r_{\text{th}}}(A) = -2/A^2$. Also it controls the size of the throat, a larger charge $A$ gives a larger wormhole throat.

The above results indicate that the presence of the phantom
scalar field sources the wormhole geometry defining  the scalar curvature and specifying  the throat of the wormhole geometry. As we will see in the next Section if the scalar curvature is generalized to a general $f({\sf R})$ function, the phantom scalar field sources the wormhole geometry.

\subsection{$f(R)$ Gravity Phantom Wormhole}

 A generalization of the Ellis Drainhole solution is to introduce a potential for the scalar field. Then we consider the full action (\ref{action}) from which we obtain the field equations (\ref{EE}) and (\ref{KG}). To solve them we consider the metric ansatz (\ref{MorrisThorne-metric}). As we discussed in Section (\ref{sect2}) the field equations resulting from action (\ref{action})  are (\Ref{KG1}), (\ref{ttrr}) and (\ref{ttuu}).
These equations constitute a system of three independent differential equations for the four unknown functions: $b(r),~ f_R(r),~ V(r),~ \phi(r)$, therefore in principle we have to fix one of the unknown functions to solve for the others. One can also see that equation (\Ref{KG1}) can be obtained by taking the covariant derivative of equation  (\Ref{EE}). We found that for a particular scalar field configuration we can obtain a rather simple solution. Therefore we fix
\begin{equation} \phi(r) = \sqrt{6} \frac{\sqrt{A}}{r}~,\label{scal} \end{equation}
where $A>0$ is a constant which has dimensions  $[L]^2$.  Then solving the equations (\Ref{ttrr}), (\Ref{ttuu}) and (\Ref{KG1}) we get
\begin{eqnarray}
b(r) &=& \frac{c_1 r^3}{A^2}+2 r~,\label{mas}\\
f_R(r) &=& \frac{A}{r^2}~, \label{fR}\\
V(r) &=& -\frac{3 c_1}{A r^2}~,\\
f(r) &=& \frac{2 A}{r^4}~,\\
R(r) &=& \frac{6 c_1}{A^2}+\frac{4}{r^2}~,\\
f(R) &=& \frac{\left(A^2 R-6 c_1\right)^2}{8 A^3}~,\\
V(\phi) &=& -\frac{c_1 \phi ^2}{2 A^2}~,
\end{eqnarray}
where $c_1$ is a constant of integration with units $[L]^2$ in order $b(r)/r$ to be dimensionless. We can see that the resulting potential is a mass term potential, hence we set: $c_1=-A^2 m^2$ and now the configurations become
\begin{eqnarray}
b(r) &=& 2 r-m^2 r^3~,\\
V(r) &=& \frac{3 A m^2}{r^2}~,\\
R(r) &=& \frac{4}{r^2}-6 m^2~, \label{R} \\
f(R) &=& \frac{1}{8} A \left(6 m^2+R\right)^2~,\\
V(\phi) &=& \frac{1}{2}m^2 \phi ^2~.
\end{eqnarray}
To understand the physical meaning of the constants entering the solution of the field equations we found, let us rewrite the action (\ref{action}) using the solution we found
\begin{equation} S = \int d^4x \sqrt{-g} \left( \frac{1}{16} A \left(6 m^2+R\right)^2 + \frac{1}{2}\partial^{\mu}\phi\partial_{\mu}\phi -\frac{1}{2}m^2 \phi ^2 \right) ~,\label{action1} \end{equation}
where the scalar field is given by (\ref{scal}). We can see that the constant $A$ which enters in the choice of the scalar field
(\ref{scal}) is giving a mass to the scalar field and in the same time modifies the Ricci scalar curvature by a non-linear correction term. On the other hand it also contributes to the size of the throat of the wormhole as can be seen in (\ref{mas}).

This solution is a generalization of the Ellis wormhole solution discussed in the previous subsection. In this solution the scalar curvature has been generalized to an arbitrary $f({\sf R})$ function  and a self-interacting potential is present. If we choose a form of the phantom scalar field $\phi(r) = \sqrt{6A}/r$ and we set $f(R)=R$ we have a system of three independent equations with two unknown functions $b(r),~V(r)$ making the system over-determined and a solution satisfying the field equations for any $r$ cannot be found.  We would also like to note that the obtained $f( \sf R)$ model resembes the Starobinski model of inflation \cite{Starobinsky:1980te}, containing a Ricci scalar term, a cosmological constant term given by the mass of the scalar field and a quadratic term of the Ricci scalar.

If one uses $f_R=1$, solving the field equations, one finds
\begin{eqnarray}
&&V(r) = c_2~,\\
&&f(r) = 2c_2 -2A^2/r^4~,\\
&&f(R) = 2c_2 + R~,
\end{eqnarray}
where $b(r),\phi(r),R(r)$ are given by (\Ref{bellis}), (\Ref{ellisscalar}), (\Ref{rellis}) respectively and $c_2$ is a constant. One can see that the scalar field now does not have any self-interactions, the solution is the Ellis wormhole and the whole action becomes
\begin{equation}S = \int d^4x \sqrt{-g} \left( \frac{2c_2 + R}{2} + \frac{1}{2}\partial^{\mu}\phi\partial_{\mu}\phi -c_2 \right)  \rightarrow S = \int d^4x \sqrt{-g} \left(\frac{R}{2} + \frac{1}{2}\partial^{\mu}\phi\partial_{\mu}\phi\right)~,\end{equation}
which  is the Ellis wormhole action as expected.

 The obtained solution is supported by the scalar field and, in particular, the integration constant  $A$. For a vanishing $A$ the solution does not exist, which is also the situation in the Ellis drainhole we discussed previously.  The $f(\sf R)$ model also satisfies the Dolgov-Kawasaki stability criterion \cite{Faraoni:2006sy, Faraoni:2008mf, Bertolami:2009cd, DeFelice:2010aj} which states that $f_{RR}>0$. For our solution we have that $f_{RR}=A/4>0$. Another desired property of $f( \sf R)$ theories is the absence of ghosts in the context of cosmological perturbations which corresponds to the condition $f_R>0$. In our case as we can see in (\ref{fR}) this condition is satisfied since $A>0$.

The Kretschmann scalar and the norm of the Weyl tensor read
\begin{eqnarray}
K =R_{\alpha \beta \gamma \delta}R^{\alpha \beta \gamma \delta} &=& -\frac{16 m^2}{r^2}+12 m^4+\frac{16}{r^4}~,\\
C_{\alpha \beta \gamma \delta}C^{\alpha \beta \gamma \delta} &=& \frac{16}{3 r^4}~.
\end{eqnarray}
We can see that the only curvature singularity is for $r \to 0$ which lies out of the range of the wormhole geometry, hence the region of interest is free of singularities.
Now we proceed  to check if our solution satisfies the afforementioned wormhole criteria.
\begin{enumerate}
    \item Condition $\cfrac{b(r)}{r}\leq 1$ reduces to $m^2r^2\geq1 $ and is satisfied for every $r$ in $[\frac{1}{m},+\infty)$.
    \item $\cfrac{b(r)}{r} = 1$ yields the throat location $r_{th} = \cfrac{1}{m}$.
    \item Solving the flaring-out condition $b'(r)<\cfrac{b(r)}{r}$ we arrive at $m^2r^2>0$ which is satisfied for every $r>0$ and $m\neq0$. Finally, it is easy to verify that $b'(r_{th})=-1$ hence the relation $b'(r_{th})\leq 1$ always holds true.
\end{enumerate}
\begin{figure}[h]
\centering
	\includegraphics[width=.40\textwidth]{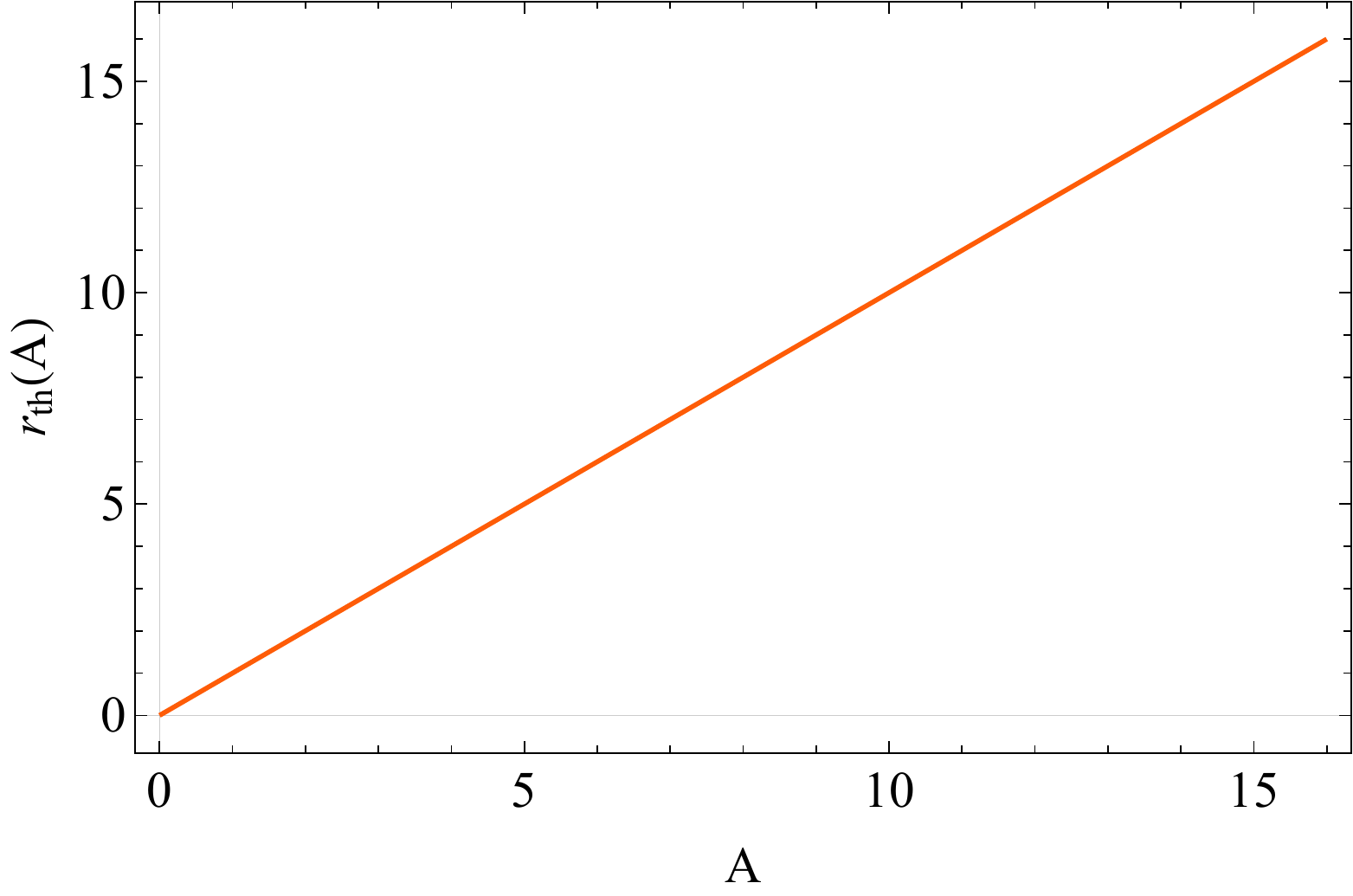}
	\includegraphics[width=.40\textwidth]{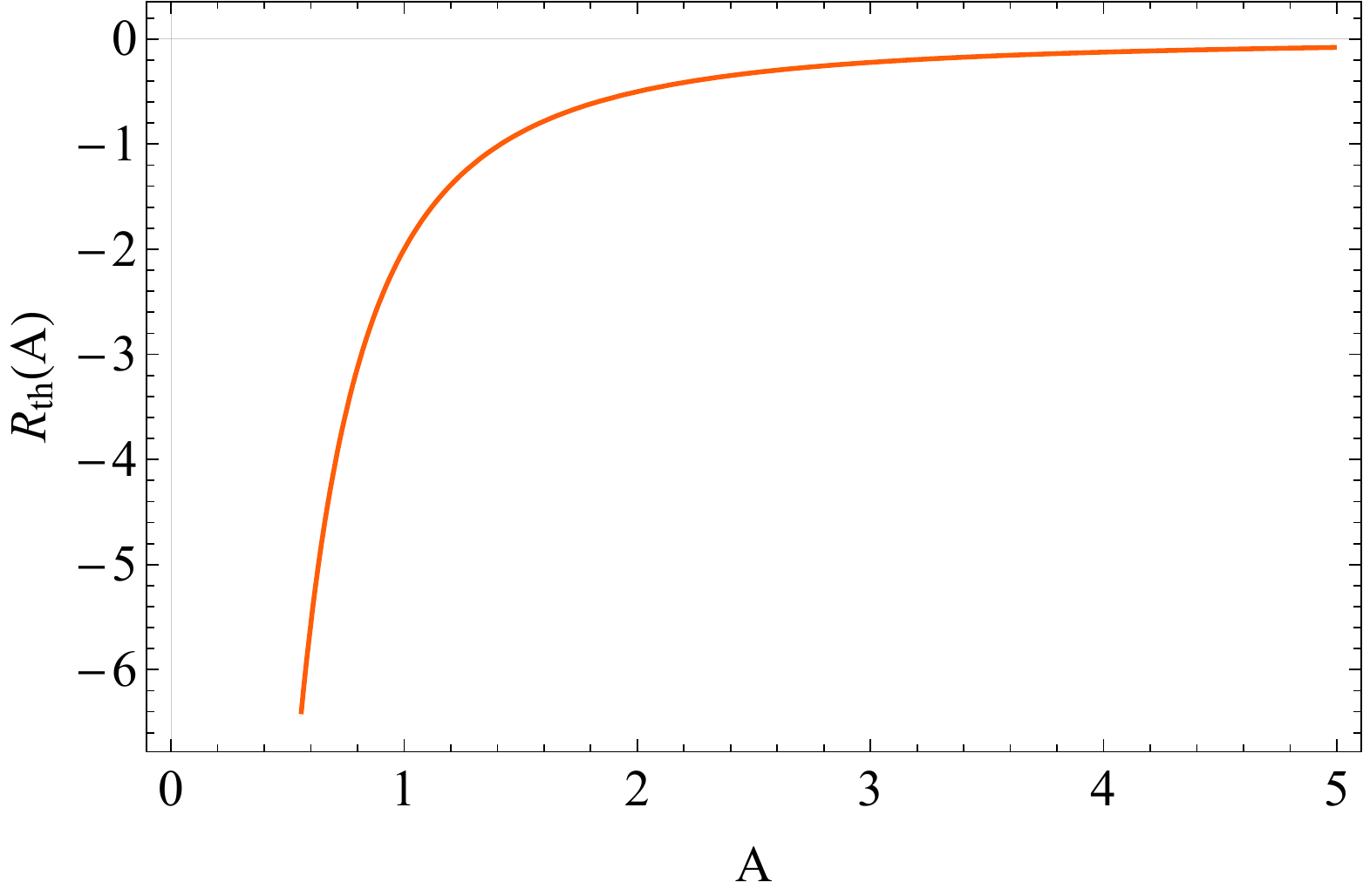}
\caption{The wormhole throat $r_{\text{th}}(A)$ and the Ricci scalar $R_{\text{th}}(A)$ at the wormhole throat as a function of the scalar constant $A$, having set $c_1=-1$.} \label{throat}
\end{figure}
In FIG. \ref{throat} we plot the throat radius $r_{\text{th}} = m^{-1} = \sqrt{-A^2/c_1}$ as a function of  $A$, where we can see their linear relation, having set $c_1=-1$. Wormholes exist only for negative $c_1$. To understand better what exactly happens at the throat of the wormhole, we also plot the Ricci scalar $R(r)$ at the throat as a function of the parameter $A$: $R_{r = r_{\text{th}}}(A) = -2m^2 = 2c_1/A^2$. A calculation of the Kretschmann scalar at the wormhole throat yields $K = 12 c_1^2/A^4$, which is in agreement with the behavior of the Ricci scalar.
We can see that for larger $A$ the Ricci scalar gets weaker at the wormhole throat, so does the Kretschmann scalar. Thus, by increasing $A$ we obtain wormholes with larger throat radii which in turn decreases the curvature in the vicinity of the throat. Hence, $A$  affects the properties of our compact object in a similar manner with Ellis's drainhole.

Additionally, we can perform a co-ordinate transformation to a chart $(t,l,\theta,\phi)$ where $l$ is the proper radial distance from the throat. In this chart the metric would be given by \eqref{metric-proper-chart}. Hence, by confronting the metrics \eqref{MorrisThorne-metric} and \eqref{metric-proper-chart} we can find the line element in this new coordinate system
\begin{eqnarray}
ds^2 &=&-C dt^2+dl^2+r(l)^2d\Omega^2~, \\
        &=&-C dt^2+dl^2+\left(\frac{e^{-lm}+e^{lm}}{2m}\right)^2d\Omega^2~,
\end{eqnarray}
where, $r(l)=\cfrac{e^{-lm}+e^{lm}}{2m} = \cfrac{1}{m}\cosh(lm)$ and $l\in (-\infty,\infty)$ covering both sides of the wormhole.
\begin{figure}[h]
\centering
	\includegraphics[width=.40\textwidth]{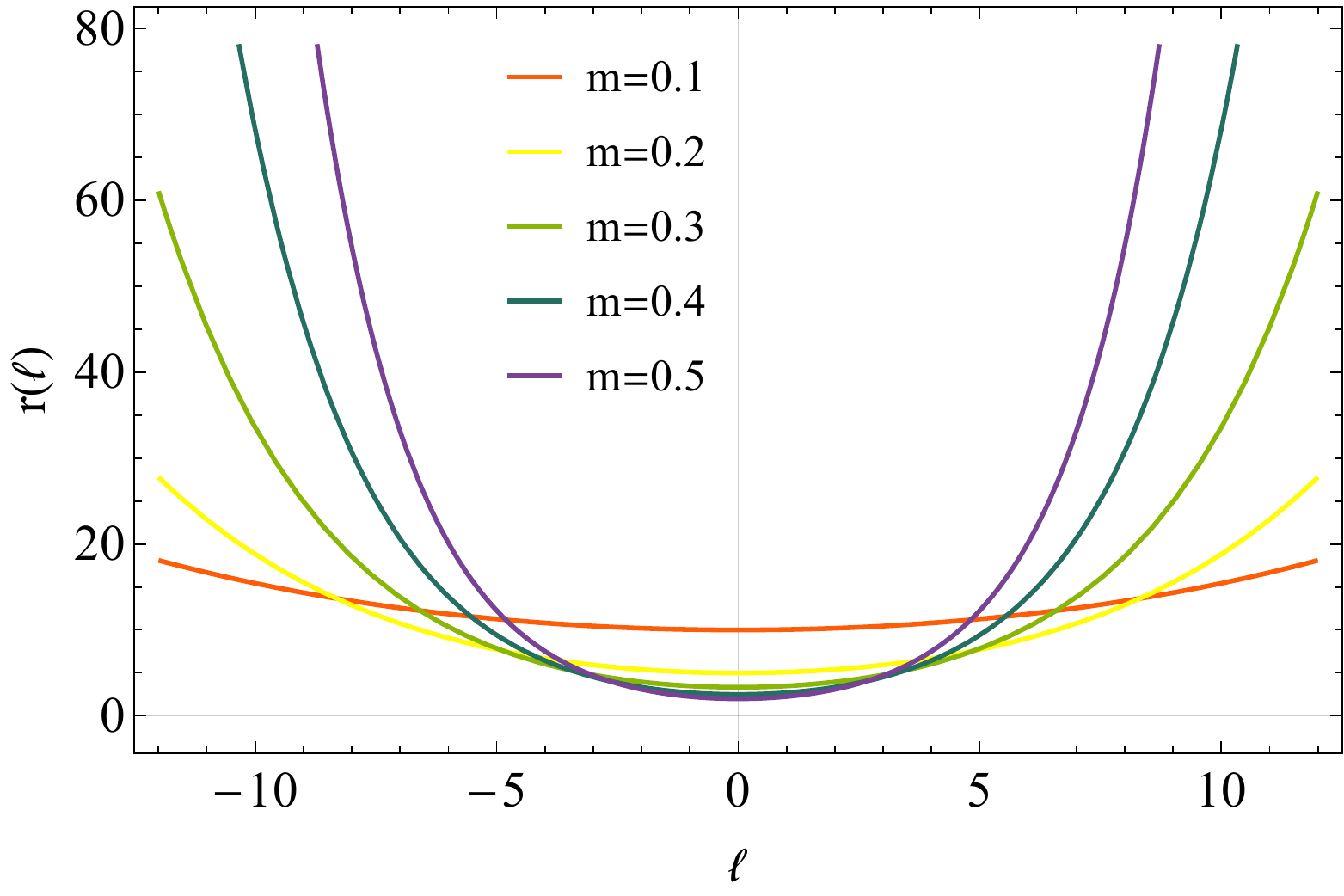}
\caption{The function $r(l)$ for different values of the mass of the scalar field.} \label{r(l)}
\end{figure}
In  FIG. \ref{r(l)} we can see the proper radial distance. The minimum represents the position of the wormhole throat.

\subsection{Energy Conditions}

Since we have obtained a wormhole configuration that satisfies the relative wormhole criteria we will briefly discuss the energy conditions and their violation. Therefore we recast Einstein equation (\Ref{EE}) in the form
\begin{equation} G_{\mu\nu} = T_{\mu\nu}^{grav} + f_{R}^{-1}T_{\mu\nu}^{\phi} = T_{\mu\nu}^{total}~, \label{EEEC} \end{equation}
where, $T_{\mu\nu}^{grav}$ is given by
\begin{equation} T_{\mu\nu}^{grav} = \frac{1}{2}g_{\mu\nu}\big(f_R^{-1}f(R) - R\big) - f_{R}^{-1} \big(g_{\mu\nu}\Box f_R - \nabla_{\mu}\nabla_{\nu}f_R\big)~ \end{equation}
and can be regared as the effective stress energy tensor that non-linear gravity generates. Now, we switch to a set of orthonormal basis vectors, the proper reference frame where the basis vectors are expressed as \cite{Morris}
\begin{equation} e_{\hat{t}} = \exp(-\Phi)e_{t}, \hspace{0.5cm} e_{\hat{r}} = (1-b/r)^{1/2}e_{r}, \hspace{0.5cm} e_{\hat{\theta}} = r^{-1}e_{\theta}, \hspace{0.5cm}  e_{\hat{\varphi}} = (r\sin\theta)^{-1}e_{\varphi}. \end{equation}
In this reference frame, we can identify, the energy density $\rho$, the radial pressure $p_r$ and the transverse pressure $p_{\theta,\varphi}$ in the following manner \cite{Visser:2003yf, Visser:1995cc}
\begin{eqnarray}
\rho &=& G_{\hat{t}\hat{t}} = T_{\hat{t}\hat{t}}^{total}~,\\
p_r &=& G_{\hat{r}\hat{r}} = T_{\hat{r}\hat{r}}^{total}~,\\
p_{\theta,\varphi} &=&  G_{\hat{\theta}\hat{\theta},\hat{\varphi}\hat{\varphi}} = T_{\hat{\theta}\hat{\theta},\hat{\varphi}\hat{\varphi}}^{total} ~.
\end{eqnarray}
Now we will discuss the violation of the energy conditions \cite{Rahaman:2006vg}. The Weak Energy Condition (WEC) implies that any time like vector $x^{\mu}$ satisfies the condition \begin{equation}T_{\mu\nu}^{total}x^{\mu}x^{\nu} \ge 0~.\end{equation}
In the afforementioned reference frame this inequality takes the form
\begin{equation} \rho>0 ~.\end{equation}
In our case we have
\begin{equation} \rho = \frac{2}{r^2}-3 m^2 >0 ~.\label{energy}\end{equation}
The WEC is satisfied for any $m\neq0$ and $0<r<\sqrt{\frac{2}{3m^2}} <\frac{1}{m}=r_{\text{th}}$. Therefore WEC is violated for any $r>r_{\text{th}}$.

The Null Energy Condition (NEC) states that
\begin{equation} T_{\mu\nu}^{total}k^{\mu}k^{\nu} \ge 0 ~,\end{equation}
for any null vector $k^{\mu}$. By continuity we expect that the WEC imples NEC. In the orthonormal frame the NEC reads
\begin{equation} \rho + p_{r} \ge 0 ~,\end{equation}
 In our case NEC gives
\begin{equation} \frac{2-3 m^2 r^2}{r^2}+m^2-\frac{2}{r^2} = -2m^2 <0~. \end{equation}
Therefore NEC is always violated.

To have a better understanding on how the energy conditions are violated, we will discuss the energy conditions of the scalar field and the modified gravity part independently. For the scalar field we will only consider the $f_R^{-1}T^{\phi}_{\mu\nu}$ part of the energy momentum tensor in equation (\Ref{EEEC}). In the afforementioned reference frame, we find that the energy density and the pressure are respectively
\begin{eqnarray}
&&\rho^{\phi} = \frac{3}{r^2}~,\\
&&p_{r}^{\phi} = \frac{3}{r^2}-6 m^2~.
\end{eqnarray}

The WEC energy condition is satisfied for the scalar field while for the NEC we have that
\begin{equation}
\rho^{\phi} + p_{r}^{\phi} = \frac{6}{r^2}-6 m^2 >0~,
\end{equation}
which is satisfied for $r<\frac{1}{m}=r_{th}$ (outside of the wormhole region) and is violated outside and on the wormhole throat.

Now for the gravitational part of the energy momentum tensor we will only examine the $T_{\mu\nu}^{grav}$ term of (\Ref{EEEC}). We find that the energy and pressure of the higher order gravity terms are respectively
\begin{eqnarray}
&&\rho^{grav} = -3 m^2-\frac{1}{r^2}~,\\
&&p_{r}^{grav} = 7 m^2-\frac{5}{r^2}~.
\end{eqnarray}
The WEC is violated by the higher order gravity terms, while NEC yields
\begin{equation}
\rho^{grav} +p_{r}^{grav} = 4m^2-\frac{6}{r^2}>0~,
\end{equation}
which holds for $r>\sqrt{\frac{3}{2}} \frac{1}{m}$ and is violated otherwise.

It is clear that if we add the energy densities of the matter and gravity part we obtain the total energy density (\Ref{energy}) and the same happens with the pressures as expected. We note here that the modified gravity part of the action is responsible for the violation of WEC, while the NEC is violated by both matter and gravity for some range of the wormhole region $r\ge r_{th}=1/m$.

\section{Conclusions}

\label{conc}

We studied wormhole solutions in a modified gravity theory in which the scalar curvature is generalized to a $f({\sf R})$ function. In this theory we considered a scalar field with negative kinetic energy, a phantom scalar field, with a self-interacting potential. Solving the field equations, choosing a
function for the phantom scalar field, we find a wormhole solution without specifying the form of the $f({\sf R})$ function. The basic properties of this solution are that the presence of the phantom scalar field influences   the scalar curvature and  the size of the throat, and by increasing the strength of the scalar field   we obtain wormholes with larger throat radii which in turn decreases the curvature in the vicinity of the throat.

Our obtained gravitational model satisfies the Dolgov-Kawasaki stability criterion, is tachyonically stable and it is free of ghosts. We calculated the NEC and the WEC and we showed that they are violated.  We also investigated the matter and gravity part of the effective energy-momentum tensor independently to see their effects on the energy conditions. We found that the scalar filed respects WEC and violated NEC, while the gravity part of the effective energy-momentum tensor violates both WEC and NEC. We also note that the $f(\sf R)$ model we obtained resembles the Starobinsky inflation model \cite{Starobinsky:1980te}.

 Our solution cannot be reduced to GR. To do that we have to  consider a more general metric ansatz such the one in \cite{Bronnikov:2018vbs} and try to find a solution to the $f({\sf R})$ field equations that reduces to the known solutions of GR \cite{Bronnikov:2018vbs, Huang:2020qmn, Huang:2019arj}. In our work, the scalar field is minimally coupled to gravity via the volume element, so one could consider a more general ansatz like a regular non-minimally coupled scalar field that will violate the energy conditions and try to find wormhole solutions. It would also be interesting to check the stability of the obtained wormhole solution.

One can fix the $f({\sf R})$ function from the beginning and then look for wormhole solutions. For example in \cite{Karakasis:2021rpn} a non-linear correction term in the scalar curvature was introduced $f({\sf R})=\sf R-2\alpha \sqrt{\sf R}$ where $\alpha$ has the dimensions of $|L|^{-1}$. In this modified gravity theory a new scale is introduced. Therefore to find  a wormhole solution that resembles the original Ellis wormhole \cite{Ellis:1973yv, Canate:2019spb} (or the $d$-dimensional generalization \cite{Rahaman:2008tq}) another scale has to be introduced to the theory to counterbalance the introduced gravitational scale. The easiest way to do that is to introduce a new scale via a self-interacting  potential for the scalar field.  So, one can consider our metric ansatz (\ref{MorrisThorne-metric}) and the Ellis scalar field (\ref{ellisscalar}) to find, at least in principle, a wormhole geometry, which will be reduced to the Ellis one, after parameterizing the constants of the solution appropriately. Wormholes with dynamical redshift function supported by a phantom scalar field might also be interesting to be studied.

 \acknowledgments
 We thank an anonymous referee for valuable comments that improved the quality of our paper.

\end{document}